\documentclass[12pt]{article}
\usepackage{epsfig,amsmath,amssymb,subfigure}
\usepackage{caption2}

   {\end{list}}%

\def\a{\alpha}
\def\b{\beta}

\def\m{\mu}

\def\r{\rho}
\def\s{\sigma}

\def\Aslash{{A\mkern-11mu/}}

\def\Dirac{{D\mkern-12mu/}}

\def\prslash{{\partial\mkern-9mu/}}

\def\prslash{{\partial\mkern-9mu/}}    %%_standard_Dirac_operator
\def\qslash{{q\mkern-8mu/}{\!}}

\def\idq{\int\!\! \frac{d^4\!q}{(2\pi)^4} \,\,}

\def\idx{\int\!\! d^4\!x}

\newcommand{\bea}{\begin{eqnarray}}
\newcommand{\eea}{\end{eqnarray}}
\newcommand{\beann}{\begin{eqnarray*}}
\newcommand{\eeann}{\end{eqnarray*}}
\newcommand{\ba}{\begin{array}}
\newcommand{\ea}{\end{array}}

\newcommand{\Tr}{\mathbf{Tr}}
\newcommand{\ST}{\star}

\def\psib{\bar{\psi}}
\def\Psib{\bar{\Psi}}
\def\g5{\gamma_{5}}

\def\Dp{{\mathcal{D}\mkern-12mu/}\,}
\def\Rs{{R\mkern-11mu/}\,}
\def\prslash {{\partial\mkern-9mu/}}  %operador Dirac

\def\Dcal{{\mathfrak{D}}}

\def\idx{\int\! d^{4}\!x\,}

 \def\psib{\bar{\psi}}
 \def\Psib{\bar{\Psi}}

\def\Aslash{{A\mkern-9mu/}}
 \def\Dirac{{D\mkern-12mu/}\,}
 \def\Dp{{\mathcal{D}\mkern-12mu/}\,}
 \def\Rs{{R\mkern-11mu/}\,}
 \def\prslash {{\partial\mkern-9mu/}}  %operador Dirac
 
\def\qslash  {{q\mkern-7mu/}}
 \def\Dcal{{\mathfrak{D}}}
 
 \def\Db {{\partial}_{\beta}}

 \def\aa {a_{\alpha}}
 \def\ab {a_{\beta}}
 \def\am {a_{\mu}}

 \def\g {\gamma}
 \def\mi {{\mu_1}}
\def\mii {{\mu_2}}
 \def\miii {{\mu_3}}
\def\miv {{\mu_4}}
 \def\a {\alpha}
\def\b {\beta}
\def\r {\rho}
 \def\s {\sigma}

 \def\Tr{\text{Tr}}
\def\tr{\text{tr}}

\def\limit{\lim_{\Lambda\rightarrow\infty}}
\def\id{{\rm{I}\!\rm{I}}}

\begin{document}
\begin{titlepage}
\rightline{FTI/UCM 80-2005}
\vglue 45pt

\begin{center}

{\Large \bf Noncommutative $SU(N)$ theories, the axial anomaly,
Fujikawa's method and the Atiyah-Singer index.}\\
\vskip 1.2 true cm
{\rm C.P. Mart\'{\i}n}\footnote{E-mail: carmelo@elbereth.fis.ucm.es}
 and C. Tamarit\footnote{E-mail: ctamarit@fis.ucm.es}
\vskip 0.3 true cm
{\it Departamento de F\'{\i}sica Te\'orica I,
Facultad de Ciencias F\'{\i}sicas\\
Universidad Complutense de Madrid,
 28040 Madrid, Spain}\\
\vskip 0.75 true cm

\vskip 0.25 true cm

{\leftskip=50pt \rightskip=50pt
\noindent
Fujikawa's method is employed to compute at first order in the noncommutative
parameter the $U(1)_A$ anomaly for  noncommutative $SU(N)$. 
We consider the most general Seiberg-Witten map which commutes with
hermiticity and complex conjugation and a noncommutative matrix parameter,
$\theta^{\mu\nu}$, which is of ``magnetic'' type. Our results for $SU(N)$  
can be readily generalized to cover the case of general nonsemisimple gauge  
groups when the symmetric Seiberg-Witten map is used. Connection with the 
Atiyah-Singer index theorem is also made.

\par }
\end{center}

\vspace{20pt}
\vfill
\end{titlepage}

%----------------------------------------------------- Paper

\setcounter{page}{2}

It is difficult to overstate the importance of the abelian chiral anomaly in
Physics~\cite{Frohlich:2000en,Adler:2004ih}. A most beautiful explanation
of the existence of this anomaly was supplied by
Fujikawa~\cite{Fujikawa:1980eg}, who showed that it comes from the lack of
invariance of the fermionic measure under chiral transformations.  Fujikawa's
method of computing anomalies also provides a way of easily exhibiting the
relationship between the abelian chiral anomaly and the Atiyah-Singer index
theorem~\cite{Fujikawa:1980eg, Alvarez-Gaume:1985ex}. The method in question
is called a nonperturbative method since no expansion in the coupling constant
is carried out.

The purpose of this note is to use Fujikawa's method to work out the abelian
chiral anomaly for noncommutative $SU(N)$ gauge theories with Dirac
fermions~\cite{Jurco:2001rq} 
up to first order in the noncommutative matrix parameter $\theta^{\mu\nu}$
and for the most general Seiberg-Witten map which is local at each order in
$\theta^{\mu\nu}$ and commutes with hermiticity and complex conjugation.
The case of noncommutive gauge theories with Dirac fermions and with
a nonsemisiple gauge group is also analysed when the theory is defined by
means of the symmetric Seiberg-Witten map~\cite{Calmet:2001na, Aschieri:2002mc}.

Let $a_{\mu}$ be an ordinary $SU(N)$ gauge field. Let $\psi$ denote an
ordinary massive Dirac fermion carrying a given representation of $SU(N)$.
Following ref.~\cite{Jurco:2001rq}, we construct the noncommutative fields
$A_{\mu}$ --the gauge field-- and $\Psi$ --the Dirac fermion-- by applying
the Seiberg-Witten map to their ordinary counterparts. As in
ref.~\cite{Aschieri:2002mc, Aschieri:2004ka}, we shall assume that
$\psi$ does not enter the Seiberg-Witten map that yields $A_{\mu}$, that
this map renders $A_{\mu}$ hermitian and that it commutes, the Seiberg-Witten map, with complex conjugation when acting on fermion fields. We shall also assume that at each order 
in the noncommutative matrix parameter $\theta^{\mu\nu}$ the Seiberg-Witten
map is local, i.e., that it is a polynomial of the fields and
their derivatives with dimensionless coefficients other than
$\theta^{\mu\nu}$. Note that if, barring
$\theta^{\mu\nu}$, we would allow for dimensionful coefficients,
such as masses, {\it etc ...}, then, the
Seiberg-Witten map would have an infinite number of terms at each order in
$\theta^{\mu\nu}$ and the theory would not be local at each order in
$\theta^{\mu\nu}$. It is not difficult to show that at first order in
$\theta^{\mu\nu}$ the most general Seiberg-Witten map that fulfils the
previous requirements reads
\begin{equation}
\begin{array}{l}
{A_\mu=a_\mu-\frac{1}{4}\{\aa,\Db\am+f_{\b\mu}\}+i\left(\kappa_2-\frac{\kappa_1}{2}\right)\,\theta^{\alpha\beta}\,\Dcal_\mu[\aa,\ab]+\kappa_3\,\theta^{\a\b}\,\Dcal_\mu f_{\a\b}+\kappa_4\,{\theta_\mu}^\b \Dcal^\nu f_{\nu\beta},}\\
{\Psi=\psi+\Big[-\frac{1}{2}\theta^{\alpha\beta}a_\a\Db+\frac{i}{4}\theta^{\a\b} \aa\,\ab+i\kappa_3\,\theta^{\a\b}\,f_{\a\b}-\left(\kappa_2-\frac{\kappa_1}{2}\right)\,\theta^{\a\b}\,[\aa,\ab]+z_1\,\theta^{\a\b}\,f_{\a\b}}\\
{\phantom{\Psi=\psi}+\frac{i}{2}\,z_2\,\theta^{\a\b}[\g_\b,\g_\r]D_\a D^\r-\frac{z_3}{2}\,\theta^{\a\b}\,[\g_\a,\g_\r]{f^\rho}_\b +{i}z_4\,\theta^{\a\b}\,\g_\a\g_\b D^2-z_5\,\theta^{\a\b}\,\g_\a\g_\b\g_\mu\g_\nu f^{\mu\nu}\Big]\psi.}\\
\end{array}
\label{SW}
\end{equation}
Hermiticity of $A_\mu$ demands $\kappa_{i}$, $i=1,...,4$, to be real numbers.
That the Seiberg-Witten map commutes with complex conjugation
--i.e.,
$\Psi[\psi,a_{\mu},\theta^{\mu\nu}]=
\Psi[\psi^{*},-a^{*}_{\mu},-\theta^{\mu\nu}]$, see refs.~\cite{Aschieri:2002mc, Aschieri:2004ka}-- leads to $z_2=z_3=z_4=z_5=0$
and restricts $z_1$ to be a real number. Notice that the terms in the
Seiberg-Witten map that go with $\kappa_4$ and $z_1$ correspond, respectively, 
to field  redefinitions of $a^{\mu}$ and $\psi$, so that their actual values have no effect on physical quantities. However,
we shall keep these parameters arbitrary and see whether they can be used to
simplify the values of the (non-physical) Green functions of the fields we
shall compute.

The action of the noncommutative $SU(N)$ theory we shall study is given by
\begin{displaymath}
    S\,=\,\idx -\frac{1}{4g^2}\mathrm{Tr}\,F^{\mu\nu}\ST F_{\mu\nu}\,+\,
\Psib\ST (i\Dirac_{\ST}-m)\Psi.
\end{displaymath}
$\mathrm{Tr}$ denotes the trace operation  on the matrix representation of
$SU(N)$ carried by $\psi$. In the previous equation,
$F^{\mu\nu}=\partial_{\mu}A_\nu-\partial_{\nu}A_\mu-i[A_\mu,A_\nu]_{\ST}$,
$\Dirac_{\ST}=\prslash-i\Aslash\ST$ and $A_{\mu}$ and $\Psi$ are given by
the Seiberg-Witten map above. $\ST$ stands for the Moyal product of
functions:$(f\ST g)(x)=f(x)\exp(\frac{i}{2}\theta^{\alpha\beta}
\overleftarrow{\partial_{\alpha}}\overrightarrow{\partial_{\beta}})g(x)$. Since we shall use Fujikawa's method
to compute the abelian anomaly, we must define the theory for the Euclidean
signature of space-time. Upon Wick rotation --we shall play it
safe~\cite{Gomis:2000zz} and
consider $\theta^{\mu\nu}$ to be of ``magnetic'' type: $\theta^{0 i}=0$--, we
obtain a theory whose action, $S_{E}$, at first order in $\theta^{\mu\nu}$
reads:
\begin{equation}
S_{E}=S_{YM}-\idx\;\psib\,({\cal K}+i{\cal M}(x))\,\psi.
\label{SEaction}
\end{equation}
$S_{YM}$ is the contribution coming from the pure noncommutative
Yang-Mills action --whose actual value will be irrelevant to us. The
differential operator ${\cal K}$ and the function ${\cal M}(x)$ are given by
\begin{equation}
\begin{array}{l}
{{\cal K}\,=\,i\Dirac\,+\,i \Rs}\\
{\Rs\,=\,(-\frac{1}{4}+2z_1) \,\theta^{\a\b}\,f_{\a\b}
\g^\mu D_\mu -\frac{1}{2}\,\theta^{\a\b}\,\g^\rho f_{\r\a}
D_\b
+ z_1\,\theta^{\a\b}\,\g^\mu\,\Dcal_\mu f_{\a\b}-
i\kappa_4\,{\theta_\mu}^\beta\,\Dcal^\nu f_{\nu\b}\, \g^\mu,}\\
{{\cal M}(x)\,=\,m\,[\,1+(-\frac{1}{4}+2z_1) \,\theta^{\a\b}\,f_{\a\b}(x)\,].}\\
\end{array}
\label{Kop}
\end{equation}
The operator $i \Rs$ is gauge covariant and formally self-adjoint and,
in ${\cal K}$, it should be understood as a perturbation of the ordinary Dirac
operator $i\Dirac$. Note that this perturbation does not destroy the pairing between positive and negative eigenvalues that occurs in the spectrum of 
$i\Dirac$. 

We shall assume that the ordinary Dirac operator has a discrete spectrum. The latter is achieved by imposing on the fields boundary conditions that
allow, by means of the stereographic projection, for the compactification of ordinary 4-dimensional Euclidean space to a 4-dimensional unit
sphere~\cite{Jackiw:1976dw, Chadha:1977mh}. In  particular, we shall assume that the ordinary gauge fields satisfy the standard boundary condition:
$a_{\mu}(x)\rightarrow  i g(x)\partial_{\mu}g^{-1}(x)+O(1/|x|^2)$
as $x\rightarrow\infty$. In keeping with the
philosophy adopted in this paper, we shall take for granted that the eigenvalues and eigenfunctions of $K$ can be computed by employing standard perturbation
theory, using $i \Rs$ as a perturbation. Thus, following
Fujikawa~\cite{Fujikawa:1980eg}, we shall use the eigenfunctions of
${\cal K}$ to define the fermionic measure of the path integral. One  
expands first the fermion fields $\psi(x)=\sum_{n}\,a_n \varphi_{n}(x)$,
$\psib(x)=\sum_{n}\,\bar{b}_n \varphi^{\dagger}_{n}(x)$, in terms of the
of a orthonormal set of eigenfunctions of  ${\cal K}$, say $\{\varphi_{n}(x)\}_n$.
Recall that $a_n$ and $\bar{b}_n$ are Grassmann variables. Then, the fermionic
measure is defined as follows
$d\psi d\psib = \prod_{n}da_n d\bar{b}_n$.

The generating functional, $Z[J^{a\mu},\omega,\bar{\omega}]$, of the complete Green functions of our theory is defined by the following path integral
\begin{equation}
Z[J^a_{\mu},\omega,\bar{\omega}]=\frac{1}{N}\int\,d\mu\quad
e^{-S_{E}+\idx\;[J^{a\mu}(x)a^a_{\mu}(x)+\bar{\omega}(x)\psi(x)+\psib(x)
\omega(x)]},
\label{pathintegral}
\end{equation}
where  $S_{E}$ is defined by eqs.~(\ref{SEaction}) and (\ref{Kop}), and
the path integral measure $d\mu$ is equal to
$[da^a_{\mu}]\prod_{n}da_n d\bar{b}_n$. $[da^a_{\mu}]$ is the measure over the
space of gauge fields and contains the Faddeev-Popov factor.
In the massless limit $S_{E}$ in eq.~(\ref{SEaction}) is invariant under the
following infinitesimal $U(1)$ rigid chiral transformations
$\delta\psi(x)=i\alpha\gamma_{5}\psi(x)$, $\delta\psib(x)=i\alpha\psib(x)\gamma_{5}$. Hence, under the infinitesimal local abelian chiral transformations
$\delta\psi(x)=i\alpha(x)\gamma_{5}\psi(x)$,
$\delta\psib(x)=i\alpha(x)\psib(x)\gamma_5$, the action $S_{E}$ undergoes the
change
\begin{displaymath}
\delta S_{E}\,=-\,\idx\; [\alpha(x)\partial_{\mu}j_{5}^{\mu}(x)
-2\alpha(x)\psib(x){\cal M}(x)\gamma_5\psi(x)].
\end{displaymath}
The current $j_{5}^{\mu}(x)$ is the $U(1)_A$ current, which is classically
conserved and is given by
\begin{equation}
j_{5}^{\mu}(x)\,=\,\psib(x)\left[\gamma^{\mu} -\left(\frac{1}{4}-2z_1\right) \,
\theta^{\a\b}\,f_{\a\b}
\g^\mu -\frac{1}{2}\,\theta^{\a\mu}\,\g^\rho f_{\r\a}\right]\gamma_5\psi(x).
\label{current}
\end{equation}
The measure of the path integral above also changes under the previous
local chiral transformations:
$d\mu\rightarrow d\mu\,[1-\idx \alpha(x) {\cal A}(x)]$. The symbol ${\cal A}(x)$
denotes the following formal expression
\begin{equation}
{\cal A}(x)\,=\, 2i \sum_{n}\varphi_{n}^{\dagger}(x)\gamma_5 \varphi_{n}(x).
\label{formalanom}
\end{equation}
These results and the fact that the path integral in
eq.~(\ref{pathintegral}) does not change under changes of $\psi$ and $\psib$,
leads to the following anomalous Ward identity
\begin{displaymath}
\ll[\partial_{\mu}j_{5}^{\mu}(x)
-2\psib(x){\cal M}(x)\gamma_5\psi(x)+i\bar{\omega}(x)
\gamma_5\psi(x)+i\psib(x)\gamma_5\omega(x)]\gg=
\ll{\cal A}(x)\gg,
\end{displaymath}
where $\ll\cdots\gg=\frac{1}{N}\int\,d\mu\;\cdots\;
e^{-S_{E}+\idx\;[J^{a\mu}(x)a^a_{\mu}(x)+\bar{\omega}(x)\psi(x)+\psib(x)
\omega(x)]}$.

As it stands in eq.~(\ref{formalanom}), ${\cal A}(x)$ is a formal object that
is in demand of  a proper definition. The latter is achieved as follows
\begin{equation}
{\cal A}(x)\,=\, 2i \lim_{\Lambda\rightarrow \infty}\;
\sum_{n}\varphi_{n}^{\dagger}(x)\gamma_5 e^{-\frac{\lambda_n^2}{\Lambda^2}} \varphi_{n}(x)=2i \lim_{\Lambda\rightarrow \infty}\;
\sum_{n}\varphi_{n}^{\dagger}(x)\gamma_5 e^{-\frac{{\cal K}^2}{\Lambda^2}} \varphi_{n}(x).
\label{reganom}
\end{equation}
$\lambda_n$ denotes a generic eigenvalue of ${\cal K}$, ${\cal K}$
being defined in eq.~(\ref{Kop}). The previous equation provides
a gauge invariant definition of ${\cal A}(x)$ obtained by using the operator that gives the dynamics of fermions in the chiral limit. Besides, the spectrum of
the operator $K$  has in common with the spectrum of $i\Dirac$ the following
paring property of the nonvanishing eigenvalues: for each nonvanishing
eigenvalue $\lambda_{n}$ with eigenfunction, say,  $\varphi_{n}(x)$, there
exists an eigenvalue $-\lambda_{n}$ with eigenfunction
$\gamma_5\varphi_{n}(x)$. That this pairing property holds is necessary to
establish a connection between of the value of ${\cal A}(x)$ and the index of
the operator ${\cal K}(1+\gamma_5)/2$. We shall come back to this issue
at the end of this paper.

By going over to a plane wave basis, expanding
the exponential  $e^{-\frac{{\cal K}^2}{\Lambda^2}}$, dropping all
contributions with more that one $\theta^{\mu\nu}$ and ignoring  terms that yield traces of the type $\tr\,\g_5=\tr\,\g_5\g^\mu\g^\nu=0$, one obtains
the following expression for the far r.h.s of eq.~(\ref{reganom}):
\begin{equation}
        \begin{array}{l}
{\mathcal{A}(x)=\mathcal{A}_{\rm ordinary}(x)+\mathcal{A}_{\theta}(x)\,+\,O(\theta^2),}\\
    {\mathcal{A}_{\rm ordinary}(x)=\sum_{k=2}^{\infty}
\lim_{\Lambda\rightarrow\infty}  2i\idq e^{-q^2}\frac{\Lambda^{2(2-k)}}{k!}\Tr\g_5
        \Dp^{2k}(\Lambda q)\id,}\\
        %%%%
 {\mathcal{A}_{\theta}(x)\!=\!\sum_{k=2}^{\infty}\sum_{l=0}^{k-1}\lim_{\Lambda\rightarrow\infty}2i\idq\! e^{-q^2}\frac{\Lambda^{2(2-k)}}{k!}\Tr\g_5
        \Dp^{2l}(\Lambda q)\left\{\Dirac(\Lambda q),\Rs(\Lambda q)\right\}\Dp^{2(k-1-l)}(\Lambda q)\id.}\\
     \end{array}
    \label{A1A2}
    \end{equation}
$\id$  denotes the identity function on $\rm{I\!R}^4$. Notice that $\Tr$ also 
denotes trace over $\gamma$ matrices, when there occur such matrices in
the expression affected by $\Tr$.  
The symbols $\Dirac(\Lambda q)$, $\Dp^{2}(\Lambda q)$ and $\Rs(\Lambda q)$ are defined,
respectively, by the following equalities:
\begin{displaymath}
\begin{array}{l}
{\Dirac(\Lambda q)=\Dirac+i\Lambda\qslash ,\quad
\Dp^{2}(\Lambda q)= D^2+2i\Lambda q\cdot D -\frac{i}{2} f_{\mu\nu}\gamma^{\mu}
\gamma^{\nu},}\\
 {\Rs(\Lambda q)=\Rs+i\left[(-\frac{1}{4}+2z_1)\,\theta^{\a\b}\,f_{\alpha\beta}\Lambda\qslash-\frac{1}{2}\theta^{\a\b}\gamma^\rho\,f_
        {\rho\alpha}\Lambda q_\beta\right].}\\
\end{array}
\end{displaymath}
$\Dirac$ and $\Rs$ are given in eq.~(\ref{Kop}).

$\mathcal{A}_{\rm ordinary}(x)$ gives, of course, the abelian anomaly in ordinary 4-dimensional  Euclidean space:
\begin{equation}
\mathcal{A}_{\rm ordinary}(x)=
\frac{i}{(4\pi)^2}\,\epsilon^{\mu\nu\rho\sigma} \Tr\, f_{\mu\nu}f_{\rho\sigma}.
\label{ordinary}
\end{equation}
Let us show next that the terms with $k$ such that $k\geq 5$ yield a vanishing contribution to $\mathcal{A}_{\theta}(x)$ in eq.~(\ref{A1A2}). Let us consider a term coming
from the expansion of
\begin{displaymath}
\begin{array}{l}
  {\Dp^{2l}(\Lambda q)\left\{\Dirac(\Lambda q),
     \Rs(\Lambda q)\right\}\Dp^{2(k-1-l)}(\Lambda
     q)}=\\
     {\left[D^2-\frac{i}{2}\gamma^\mu\gamma^\nu f_{\mu\nu}+2i\Lambda
     q\cdot D\right]^{l}\left\{\Dirac(\Lambda q),
     \Rs(\Lambda q)\right\}\left[D^2-\frac{i}{2}\gamma^\mu\gamma^\nu f_{\mu\nu}+2i\Lambda
     q\cdot D\right]^{k-1-l}},
\end{array}
\end{displaymath}
which contains $a$, $b$ and $c$ factors of type $D^2$,
$\gamma^\mu\gamma^\nu f_{\mu\nu}$ and $2i\Lambda q\cdot D$, respectively.
Since $\left\{\Dirac(\Lambda q),\Rs(\Lambda q)\right\}$ supplies two $\gamma$
matrices to the term in question, we conclude that the trace over the Dirac
matrices will vanish unless $2b+2 \geq 4$, i.e., unless $b\geq 1$. Now, notice
that $a+b+c=k-1$, so that $c$ is bounded from above as follows:
$c\leq c_{\max}=k-1-b$. Hence, the highest power of $\Lambda$ that occurs in the term that we are analysing is $c_{max}+2=k-b+1$. Next, this term is to be
multiplied by $\Lambda^{2(2-k)}$, so, for $k>2$, it will not survive in the
large $\Lambda$ limit if $k-b+1< 2(k-2)$. This inequality and the constraint
$b\geq 1$, leads to $k>4$. We thus conclude that
 \begin{equation}
        \mathcal{A}_{\theta}= T_1+T_2+T_3,
\label{ateta}
    \end{equation}
where $T_1$, $T_2$ and $T_3$ correspond, respectively, to the contributions
to $\mathcal{A}_{\theta}(x)$ --see eq.~(\ref{A1A2})-- with $k=2$, $3$ and $4$:
\begin{equation}
        \begin{array}{l}
        {T_1=\sum_{l=0}^{1}\limit 2i\idq e^{-q^2}\frac{1}{2}\Tr\g_5
        \Dp^{2l}(\Lambda q)\left\{\Dirac(\Lambda q),\Rs(\Lambda q)\right\}\Dp^{2(1-l)}(\Lambda q)\id,}\\
     %%%%%
        { T_2=\sum_{l=0}^{2}\limit 2i\idq e^{-q^2}\frac{1}{3!\Lambda^2}\Tr\g_5
        \Dp^{2l}(\Lambda q)\left\{\Dirac(\Lambda q),\Rs(\Lambda q)\right\}\Dp^{2(2-l)}(\Lambda q)\id,}\\
     %%%%%%
         {T_3=\sum_{l=0}^{3}\limit 2i\idq e^{-q^2}\frac{1}{4!\Lambda^4}\Tr\g_5
        \Dp^{2l}(\Lambda q)\left\{\Dirac(\Lambda q),\Rs(\Lambda q)\right\}\Dp^{2(3-l)}(\Lambda q)\id.}
    \end{array}
    \label{t1t2t3}
    \end{equation}
To carry out the computation of $T_1$, $T_2$ and $T_3$, we shall need the
expansion of $\left\{\Dirac(\Lambda q),\Rs(\Lambda q)\right\}$ in powers of
$\Lambda$:
\begin{equation}
    \begin{array}{l}
    {\left\{\Dirac(\Lambda q),\Rs(\Lambda q)\right\}=\g^\mu\g^\nu S_{\mu\nu}= \g^\mu\g^\nu(S_{\mu\nu}|_{\Lambda^0}+S_{\mu\nu}|_{\Lambda^1}+S_{\mu\nu}|_{\Lambda^2}),}\\
    {S_{\mu\nu}|_{\Lambda^0}=d_1\theta^{\a\b}\left(\Dcal_\mu f_{\a\b}D_\nu+
2 f_{\a\b}D_\mu D_\nu\right)-\frac{1}{2}\,\theta^{\a\b}
\left(\Dcal_\mu f_{\nu\a} D_\beta+f_{\nu\a}D_\mu D_\b+f_{\mu\a}D_\b
D_\nu\right),}\\
    {\phantom{S_{\mu\nu}|_{\Lambda^0}}+z_1\,\theta^{\a\b}
(\Dcal_\mu\Dcal_\nu f_{\a\b}+\Dcal_{\{\nu} f_{\a\b}D_{\mu\}^{o}})
-i\kappa_4\,({\theta_\nu}^\b\Dcal_\mu\Dcal^\a f_{\a\b}+
{\theta_{\{\nu}}^\b\Dcal^\a f_{\a\b}D_{\mu\}^{o}}),}\\
    %%%
    {S_{\mu\nu}|_{\Lambda^1}=i\Lambda (q_{\{\mu} R_{\nu\}^{o}})+
i d_1 \Lambda\,\theta^{\a\b}(q_\nu\Dcal_\mu f_{\a\b}+f_{\a\b}q_{\{\mu}D_{\nu\}^{o}})
    -\frac{i}{2}\Lambda\,\theta^{\a\b}q_\beta(\Dcal_\mu f_{\nu\a}+
f_{\{\mu\a}D_{\nu\}^{o}})}\\
    {S_{\mu\nu}|_{\Lambda^2}=-\Lambda^2\,\theta^{\a\b}\,
q_{\{\mu}\left(d_1 f_{\a\b}q_{\nu\}^{o}}-\frac{1}{2}
f_{\nu\}^{o}\a}q_\b\right).}
    \end{array}
\label{DR+RD}
\end{equation}
$\{\}^{o}$ indicates that only the indices $\mu$ and $\nu$ are symmetrized.    
$d_1=-1/4+z_1$.

Let us work out $T_1$ in eq.~(\ref{t1t2t3}). Using the fact that
$\tr\, \gamma_5=\tr\, \gamma_5\gamma_{\mu}\gamma_{\nu}=0$, one concludes that
\begin{displaymath}
T_1=\limit\!\! 2i\!\idq\! e^{-q^2}\frac{1}{2}\Tr\g_5
   \left\{-\frac{i}{2}\g^\mu\g^\nu f_{\mu\nu}\left\{\Dirac(\Lambda q),
\Rs(\Lambda q)\right\}{\id}-\frac{i}{2}\left\{\Dirac(\Lambda q),
\Rs(\Lambda q)\right\}\g^\mu\g^\nu f_{\mu\nu}\right\}.
\end{displaymath}
Substituting in the previous equation the results in eq.~(\ref{DR+RD}), one 
shows that the contribution coming from $S_{\mu\nu}|_{\Lambda^1}$ vanishes 
upon integration over $q$ and that $S_{\mu\nu}|_{\Lambda^2}$ yields a 
vanishing contribution since 
$S_{\mu\nu}|_{\Lambda^2}$ is symmetric in $\mu$ and $\nu$. Then, the
computation of the integral and traces on the r.h.s. of the previous equation
leads to
\begin{equation}
\begin{array}{l}
    {T_1=-\frac{2}{(4\pi)^2}\,\epsilon^{\mi\mii\miii\miv}\Tr(f_{\mi\mii}
S_{\miii\miv}|_{\Lambda^0}+S_{\mi\mii}|_{\Lambda^0}f_{\miii\miv})}\\
{\phantom{T_1}=\frac{i}{8\pi^2}\,\theta^{\a\b}\,\epsilon^{\mi\mii\miii\miv}
\,\Tr\,(2 d_1 f_{\a\b}f_{\mi\mii}f_{\miii\miv}-f_{\mii\a}
f_{\mi\b}f_{\miii\miv})}\\
    {\phantom{T_1=}+\frac{1}{16\pi^2}\,\theta^{\a\b}\,
\epsilon^{\mi\mii\miii\miv}\,\Tr\,\big[f_{\mi\mii}(\Dcal_\miii f_{\miv\a}
D_\b \id-2 d_1 \Dcal_\miii f_{\a\b} D_{\mu_4} \id)+\Dcal_\mi f_{\mii\a}
f_{\miii\miv}D_\b \id}\\
    {\phantom{T_1=}-2 d_1 \Dcal_\mi f_{\a\b}f_{\miii\miv}
D_\mii \id\big]+\frac{i\kappa_4}{4\pi^2}\,{\theta_\miv}^\b\,
\epsilon^{\mi\mii\miii\miv}\,\Tr\, \Dcal_\miii(f_{\mi\mii}
\Dcal^\nu f_{\nu\b}),}
    \end{array}
    \label{T1}
\end{equation}
where $\id$ is the unit function on $\rm{I\!R}^4$ and $d_1=-1/4+2z_1$.

To calculate  $T_2$ in eq.~(\ref{t1t2t3}) will be shall express it as the
sum of two terms, say,  $T_2^{(6\gamma)}$ and $T_2^{(4\gamma)}$, which involve the
computation of the trace over six and four $\gamma$ matrices, respectively:
\begin{equation}
\begin{array}{l}
{T_2\,=\,T_2^{(6\gamma)}\,+\,T_2^{(4\gamma)},}\\
{T_2^{(6\gamma)}=\sum_{l=0}^{2}\limit \left(-\frac{i}{2}\right)\idq e^{-q^2}\frac{1}{(3)!\Lambda^2}\Tr\g_5
       [(\g^\r\g^\s f_{\r\s})^l\g^\mu\g^\nu S_{\mu\nu}(\g^\kappa\g^\tau f_{\kappa\tau})^{(2-l)}]\id}\\
       { T_2^{(4\gamma)}=\limit\idq
       e^{-q^2}\frac{1}{(3)!\Lambda^2}\Tr\g_5
       [\{D^2\g^\r\g^\s f_{\r\s}+\g^\r\g^\s f_{\r\s}D^2,\g^\mu\g^\nu S_{\mu\nu}\}}\\
       {\phantom{T_2^{(4\gamma)}=}+D^2\g^\mu\g^\nu S_{\mu\nu}\g^\r\g^\s f_{\r\s}+\g^\r\g^\s f_{\r\s}\g^\mu\g^\nu S_{\mu\nu}D^2]\id}\\
       {\phantom{T_2^{(ii)}=}+\limit\idq
       e^{-q^2}\frac{2i}{(3)!\Lambda}\Tr\g_5
       [\{q\cdot D\g^\r\g^\s f_{\r\s}+\g^\r\g^\s f_{\r\s}q\cdot D,\g^\mu\g^\nu S_{\mu\nu}\}}\\
       {\phantom{T_2^{(ii)}=}+q\cdot D\g^\mu\g^\nu S_{\mu\nu}\g^\r\g^\s f_{\r\s}+\g^\r\g^\s f_{\r\s}\g^\mu\g^\nu S_{\mu\nu}q\cdot D\,]\id}.\\
\end{array}
\label{T26g-4g}
\end{equation}
In the limit $\Lambda\rightarrow\infty$, only the piece 
$S_{\mu\nu}|_{\Lambda^2}$ of $S_{\mu\nu}$ --see eq.~(\ref{DR+RD})-- contributes to $T_2^{(6\gamma)}$.
The computation of the corresponding integrals and some algebra yields
\begin{equation}
    T_2^{(6\gamma)}=\frac{i}{2(4\pi)^2}\,\theta^{\a\b}\,(-8d_1-1)\,\epsilon^{\mi\mii\miii\miv}\,\Tr\,f_{\a\b}f_{\mi\mii}f_{\miii\miv}.
\label{T26g}
\end{equation}
Since $\tr\, \gamma_5\gamma^{\rho}\gamma^{\sigma}\gamma^{\mu}\gamma^{\nu}\sim
\epsilon^{\rho\sigma\mu\nu}$, one concludes that only the antisymmetric
part of $S_{\mu\nu}$ in eq.~(\ref{DR+RD}) is relevant to the computation of
 $T_2^{(4\gamma)}$ in eq.~(\ref{T26g-4g}). Then, in the large $\Lambda$ limit
we have
\begin{displaymath}
\begin{array}{l}
    {T_2^{(4\gamma)}=\idq
       e^{-q^2}\frac{i}{(3)!}\Tr\g_5
       [\{q\cdot D\g^\r\g^\s f_{\r\s}+\g^\r\g^\s f_{\r\s}q\cdot D,
\g^\mu\g^\nu S_{[\mu\nu]}|_{\Lambda^1}\}}\\
       {\phantom{T_2^{(i)}=}+q\cdot D\g^\mu\g^\nu S_{\mu\nu}|_{\Lambda^1}
\g^\r\g^\s f_{\r\s}+\g^\r\g^\s f_{\r\s}\g^\mu\g^\nu S_{[\mu\nu]}|_{\Lambda^1}q
\cdot D]\id.}\\
\end{array}
\end{displaymath}
$S_{[\mu\nu]}|_{\Lambda^1}$ is the antisymmetric part of
$S_{\mu\nu}|_{\Lambda^1}$ in eq.~(\ref{DR+RD}). By substituting in the previous equation  the necessary integrals, and after some  algebra, one obtains the following result:
\begin{equation}
\begin{array}{l}
    {T_2^{(4\gamma)}=-\frac{1}{16\pi^2}\theta^{\a\b}\,\epsilon^{\mi\mii\miii\miv}\,\Tr\,
[f_{\mi\mii}\Dcal_\miii f_{\miv\a}D_\b \id+\Dcal_\mi f_{\mii\a}f_{\miii\miv}D_\b \id-2d_1 (f_{\mi\mii}\Dcal_\miii f_{\a\b}D_\miv\id}\\
{\phantom{T_2^{(4\gamma)}=-\frac{1}{16\pi^2}\theta^{\a\b}\,\epsilon^{\mi\mii\miii\miv}\,\Tr\,
[}+\Dcal_\mi f_{\a\b}f_{\miii\miv}D_\mii\id)].}
\end{array}
\label{T24g}
\end{equation}

From the definition of $T_3$ in eq.~(\ref{t1t2t3}), one readily learns that
there are contributions to it involving $8$, $6$ and $4$ $\gamma^{\mu}$ matrices.
The contributions with  $8$ and $6$ $\gamma^{\mu}$ matrices vanish in the large $\Lambda$ limit as $\Lambda^{-2}$ and $\Lambda^{-1}$, respectively.
The contributions with $4$ $\gamma^{\mu}$ matrices also go away as
$\Lambda\rightarrow\infty$, since in this limit they are proportional
$1/\Lambda^2\,\epsilon^{\rho\sigma\mu\nu}S_{\mu\nu}|_{\Lambda^2}$ and
$S_{\mu\nu}|_{\Lambda^2}$ is symmetric in its indices $\mu$ and $\nu$.
In conclusion:
\begin{displaymath}
T_3\,=\,0.
\end{displaymath}

Substituting the previous equation and eq.~(\ref{T24g}), (\ref{T26g})
and (\ref{T1}) in eq.~(\ref{ateta}), one obtains the following result:
\begin{equation}
\begin{array}{l}
{{\cal A}_{\theta}(x)=T_1+T_2^{(6\gamma)}+T_2^{(4\gamma)}+T_3
=-\frac{i}{32\pi^2}\,\theta^{\a\b}\,\epsilon^{\mi\mii\miii\miv}\,
\Tr\,[f_{\a\b}f_{\mi\mii}f_{\miii\miv}+4 f_{\a\miii}f_{\miv\b}f_{\mi\mii}]}\\
    {\phantom{{\cal A}_{\theta}(x)=T_1+T_2^{(6\gamma)}+T_2^{(4\gamma)}+T_3=}+\frac{i\kappa_4}{4\pi^2}\,{\theta_\miv}^\beta\,\epsilon^{\mi\mii\miii\miv}\,\Tr\,\Dcal_\miii(f_{\mi\mii}\Dcal^\nu f_{\nu\b})}\\
{\phantom{{\cal A}_{\theta}(x)=T_1+T_2^{(6\gamma)}+T_2^{(4\gamma)}+T_3}=
\partial_{\mu}\big(
\frac{i\kappa_4}{4\pi^2}\,{\theta_\miii}^\beta\,\epsilon^{\mu\mi\mii\miii}\,\Tr\,f_{\mi\mii}\Dcal^\nu f_{\nu\b}\big).}\\
\end{array}
\label{atetares}
\end{equation}
The identity
$\Tr\,\theta^{\a\b}\,\epsilon^{\mi\mii\miii\miv}[f_{\a\b}f_{\mi\mii}f_{\miii\miv}+4 f_{\a\miii}f_{\miv\b}f_{\mi\mii}]=0$ has been used to get the second equality in the previous expression.

In view of eq.~(\ref{atetares}), one concludes that there is no anomalous contribution at first order in $\theta^{\mu\nu}$. Notice that one can always set $\kappa_4=0$
and that even in the event that one insisted in having a nonvanishing
$\kappa_4$, the contribution to ${\cal A}_{\theta}$ can be absorbed by performing the following finite and gauge invariant renormalization of the current
$j^{\mu}_5$ in eq.~(\ref{current}): $j^{\mu}_{5\,\rm{ren}}=j^{\mu}_5
-\frac{i\kappa_4}{4\pi^2}\,{\theta_\miii}^\beta\,\epsilon^{\mu\mi\mii\miii}\,\Tr\,f_{\mi\mii}\Dcal^\nu f_{\nu\b}.$

Let us choose $SU(N)$, with $N>2$, as our ordinary gauge group. That
${\cal A}(x)$ in eq.~(\ref{A1A2}) be  equal to
${\cal A}(x)_{\rm ordinary}$ up to first order in $\theta^{\mu\nu}$ is a
highly nontrivial result. Indeed, ${\cal A}_{\theta}(x)$ being proportional
to a truly anomalous term like $\Tr\,\theta^{\a\b}\,\epsilon^{\mi\mii\miii\miv}[f_{\a\b}f_{\mi\mii}f_{\miii\miv}]$ is  consistent with power
counting and gauge invariance. And yet, as shown in eq.~(\ref{atetares}) all contributions of this type cancel each other. Why? One may answer this question
by establishing the connection between the abelian anomaly ${\cal A}(x)$ and
the index of ${\cal K}(1+\gamma_5)/2$, ${\cal K}$ being defined in
eq.~(\ref{Kop}). But first let us exhibit some properties of
${\cal P}(x)=\Tr\,\theta^{\a\b}\,\epsilon^{\mi\mii\miii\miv}[f_{\a\b}f_{\mi\mii}f_{\miii\miv}(x)]$.

The first property we want to display is that for $SU(2)$, ${\cal P}(x)=0$.
The second property is that for $SU(N)$, with $N>2$, ${\cal P}(x)$ cannot be
expressed as $\partial_{\mu}\,X^{\mu}$, $X^{\mu}$ being a gauge invariant
polynomial of the gauge field and its derivatives. This is why we
called  ${\cal P}(x)$ a truly  anomalous contribution for $SU(N)$, $N>2$.
That ${\cal P}(x)$ possesses this property can be shown as follows. If there
exist such an $X^{\mu}$, it would be a polynomial on the field $a^{a}_{\mu}$
and its derivatives such that  $s_0\,X^{\mu}|_{aaa}=0$. $s_0$ is the free
BRS operator --$s_0\, a^{a}_{\mu}=\partial_{\mu}c^a$-- and $X^{\mu}|_{aaa}$
is the contribution to $X^{\mu}$ which has 3 fields $a^{a}_{\mu}$ and 2
partial derivatives. Now, it has been shown in ref.~\cite{Brandt:1989rd}
that the cohomology of $s_0$ over the space of polynomials of $a^{a}_{\mu}$ and
its derivatives is constituted by polynomials of
$f_{0\,\mu\nu}^{a}=
\partial_{\mu}a^{a}_{\nu}-\partial_{\nu}a^{a}_{\mu}$ and its
derivatives. Hence, $X^{\mu}|_{aaa}=0$, for it cannot expressed as
a polynomial of $f_{0\,\mu\nu}^{a}$ and its derivatives: we are
one derivative short in $X^{\mu}|_{aaa}$. $X^{\mu}|_{aaa}=0$
implies that $X^{\mu}$ does not exist. The third property is that
$\idx\; {\cal P}(x)$ does not necessarily vanish for fields with
well-defined Pontrjagin number. For instance, in the $SU(3)$ case,
$a^{a}_{\mu}=a^{({\rm BPST})\,a}_{\mu}+\delta^{a8}\,b_{\mu}$, with
$a^{({\rm BPST})\,a}_{\mu}$ being standard embedding of the BPST
$SU(2)$ instanton into $SU(3)$ and $b_{\mu}$ being a 4-dimensional
vector field with components $b_\mu=\frac{\omega_{\mu\nu}x_\nu
\rho^{2(n-1)}}{(r^2+\rho^2)^n},\,\,
n>1,\,\omega_{\mu\nu}=-\omega_{\nu\mu},\,\,\omega_{\mu\nu}=\text{Sign}(\nu-\mu)(\mu\cdot\nu)
\,\text{ for $\mu\neq\nu$}$, yields 
$\idx\;{\cal P}(x)\neq 0$ and has Pontrjagin number equal to 1. To symplify the computation choose a $\theta^{\mu\nu}$ with $\theta^{12}=-\theta^{21}$ as its  
only nonvanishing components.

Let us now establish the connection between the abelian axial anomaly and
the index of ${\cal K}(1+\gamma_5)/2$. Using eq.~(\ref{reganom}), one
readily shows that
\begin{displaymath}
-\frac{i}{2}\,\idx\;{\cal A}(x)=\idx\sum_{n=1}^{n_+}\idx\,
\varphi_{n\,+}^{\dagger}(x)\varphi_{n\,+}(x)-
\idx\sum_{n=1}^{n_-}\idx\,
\varphi_{n\,-}^{\dagger}(x)\varphi_{n\,-}(x)=n_{+}-n_{-}.
\end{displaymath}
$n_{+}$ and $n_{-}$ are respectively the number of positive and negative
chirality zero modes of ${\cal K}$ in eq.~(\ref{Kop}). Of course,
$n_{+}=\text{ dim Ker\;}{\cal K}(1+\gamma_5)/2$ and
$n_{-}=\text{ dim Ker\;}{\cal K}^{\dagger}(1-\gamma_5)/2$. Hence, the
abelian anomaly is given by the index of ${\cal K}(1+\gamma_5)/2$:
\begin{equation}
-\frac{i}{2}\,\idx\;{\cal A}(x)\,=\,\text{ index }{\cal K}(1+\gamma_5)/2.
\label{anomindex}
\end{equation}
Now, we have assumed that the operator ${\cal K}$ differs from
the Dirac operator $i\Dirac$ in a ``infinitesimally small'' --otherwise our expansions in $\theta^{\mu\nu}$ would not make much sense-- operator
$i\Rs$ that is hermitian and such that
$\gamma_5\,{\cal K}=-{\cal K}\gamma_5$. Then, one would hope~\cite{Green:1987mn}
that
\begin{equation}
\text{ index }{\cal K}(1+\gamma_5)/2\,=\,\text{ index }i\Dirac(1+\gamma_5)/2
\,=\,\frac{1}{32\pi^2}\,\idx\;\epsilon^{\mu\nu\rho\sigma}\,\Tr\;
f_{\mu\nu}(x)f_{\rho\sigma}(x).
\label{ASindex}
\end{equation}
A by-product of our calculations is that the previous equation indeed holds as
far as we have computed. Notice that if ${\cal A}_{\theta}$ in eq.~(\ref{A1A2}) had received a 
contribution like ${\cal P}(x)=\Tr\,\theta^{\a\b}\,\epsilon^{\mi\mii\miii\miv}[f_{\a\b}f_{\mi\mii}f_{\miii\miv}(x)]$, then, in view of the discussion in the previous paragraph and eq.~(\ref{anomindex}), we would have concluded that
the first equality in eq.~(\ref{ASindex}) would not be correct. Obviously,
this analysis  can be extended and conjecture that at any order in $\theta$  the
abelian anomaly for noncommutative $SU(N)$ is saturated by the ordinary abelian
anomaly. This conjecture is further supported by the second order in
$\theta^{\mu\nu}$ Feynman diagram calculations carried out in
ref.~\cite{Martin:2005gt}.

Finally, our results can be readily extended to the case of noncommutative
gauge theories with a nonsemisimple gauge group, when the noncommutative theory is constructed  by using the symmetric 
form of the Seiberg-Witten map as defined in ref.~\cite{Aschieri:2002mc}.
In this case the Seiberg-Witten map is the same as the map displayed in
eq.~(\ref{SW}) by now $a_{\mu}$ is given by 
$    a_\m = \sum_{k=1}^s g_k\, (a_\mu^k)^a\,(T^k)^a
         + \sum_{l=s+1}^N g_l\,a_\mu^l\,T^l$ 
and the spinor $\psi$ denotes a hypermultiplet carrying a given
representation of the nonsemisimple gauge group. 
$a_\m^k$, $g_k$ and $a_\mu^l$, $g_l$ are the ordinary gauge field
and coupling constants associated,
respectively, to each simple and $U(1)$ factor of the nonsemisimple group. 
The reader is referred to  
 ref.~\cite{Brandt:2003fx} for further details on the notation. It is clear
that eqs.~(\ref{ordinary}) and (\ref{atetares}) will also be valid in the
nonsemisimple case provided $a_\m$ is defined as in the previous equation.

It is a very interesting and open question to obtain the results presented
in this paper by using the heat kernel expansion~\cite{Vassilevich:2003xt}
due to its relevance in the mathematically rigorous proof of index theorems.

\section*{Acknowledgments}

This work has been  financially supported in part by MEC through grant
BFM2002-00950. The work of C. Tamarit has  also received financial support
from MEC trough FPU grant AP2003-4034.

\end{document}